# Observation of inverse spin Hall effect in bismuth selenide

Praveen Deorani,[1] Jaesung Son,[1] Karan Banerjee,[1] Nikesh Koirala,[2] Matthew Brahlek,[2] Seongshik Oh,[2] and Hyunsoo Yang[1,*]

[1]*Department of Electrical and Computer Engineering, National University of Singapore, 117576, Singapore*

[2]*Department of Physics & Astronomy, Rutgers Center for Emergent Materials, Institute for Advanced Materials, Devices and Nanotechnology, The State University of New Jersey, New Jersey 08854, USA*

Bismuth Selenide ($Bi_2Se_3$) is a topological insulator exhibiting helical spin polarization and strong spin-orbit coupling. The spin-orbit coupling links the charge current to spin current via the spin Hall effect (SHE). We demonstrate a $Bi_2Se_3$ spin detector by injecting the pure spin current from a magnetic permalloy layer to a $Bi_2Se_3$ thin film and detect the inverse SHE in $Bi_2Se_3$. The spin Hall angle of $Bi_2Se_3$ is found to be $0.0093 \pm 0.0013$ and the spin diffusion length in $Bi_2Se_3$ to be $6.2 \pm 0.15$ nm at room temperature. Our results suggest that topological insulators with strong spin-orbit coupling can be used in functional spintronic devices.

In a solid, the spin of an electron is coupled to its momentum via spin-orbit interaction. This coupling is manifested in the direct and inverse spin Hall effect (SHE and ISHE) [1-7], which are the processes of inter-conversion of charge and spin currents. The SHE and ISHE have been observed in many materials, including metals and semiconductors. The generation and detection of spin currents is of key importance in spintronics [8-10]. In particular, the ISHE, which converts the spin current into a charge electromotive force (emf), can be used as a powerful detection mechanism for spin currents.

Topological insulators (TI) are a new quantum state of matter where non trivial topology of bulk bands are manifested in gapless spin polarized surface states [11-13]. To date, most of the experiments in TIs have focused on the edge or surface state of these materials. In 2D TIs, edge states have been confirmed by conductance quantization in the insulating regime [14], and the helical spin polarization of these states has been detected by transport experiments like the quantum spin Hall effect [15]. In 3D TIs, there have been many studies which employ angle-resolved photoemission spectroscopy (ARPES) to observe the Dirac cones and confirm its non-degenerate nature. Besides that, the spin-resolved ARPES has been conducted to confirm the helical spin polarization in surface states in 3D TIs [16, 17]. The surface state conduction has been confirmed by sample thickness dependence of transport properties [18-20], and by quantum oscillations from the surface state [18, 21-23]. However, spin transport is little studied in TIs because of the difficulties such as impedance mismatch [24] and an expected small spin diffusion length in TIs due to strong spin-orbit coupling [25]. Recently, spin-orbit effects have been observed in TIs, reporting a wide range of spin Hall angle. [26-28]

In this work we report the observation of ISHE in the prototypical 3D TI, $Bi_2Se_3$, and demonstrate its feasibility as a spin current detector. We utilize spin pumping from a permalloy (Py) layer to generate spin currents into the $Bi_2Se_3$ layer, and measure the induced emf signal [6, 29-32]. Spin pumping is free from the impedance mismatch problem [33] and provides high density spin current injection into the $Bi_2Se_3$ layer. From thickness dependent measurements, the spin Hall angle ($\theta_{sh}$) and spin diffusion length ($\lambda_{sf}$) in $Bi_2Se_3$ thin films are obtained to be $0.0093 \pm 0.0013$ and $6.22 \pm 0.15$ nm, respectively, at room temperature (RT). Both the $\theta_{sh}$ and $\lambda_{sf}$ are observed to increase as the temperature decreases.

$Bi_2Se_3$ thin films were grown on $Al_2O_3(0001)$ substrates using custom designed SVTA MOSV-2 MBE system with base pressure less than $3\times10^{-10}$ Torr. 99.999% pure bismuth and selenium sources were placed in Knudsen cells and evaporated to provide stable fluxes. An Inficon BDS-250 XTC/3 Quartz crystal microbalance was used to measure the flux of individual elements. kSA 400 reflection high energy electron diffraction (RHEED) measurement system was used to monitor the crystal quality of $Bi_2Se_3$ thin films during growth. $Al_2O_3(0001)$ substrates were cleaned for 5 minutes using *ex situ* UV/ozone cleaner to remove organic contaminants on the surface and were immediately loaded into the MBE growth chamber afterwards. In the growth chamber, the substrates were heated to 750 ºC for 10 minutes in $1\times10^{-6}$ Torr oxygen environment to remove any excess contaminants. The substrates were then cooled to growth temperature of 135 ºC, while the sources were heated to desired temperature of 600 ºC and 186 ºC for Bi and Se, respectively. We employ a two step growth procedure for $Bi_2Se_3$ thin film growth as reported previously [34]. A seed layer of 3 quintuple layer

Bi$_2$Se$_3$ was grown at 135 ºC followed by annealing to 220 ºC for 10 minutes before depositing rest of the film at 220 ºC. Highly crystalline Bi$_2$Se$_3$ films were obtained as indicated by sharp streaks and associated Kikuchi lines in the RHEED pattern. The carrier concentration of the Bi$_2$Se$_3$ film was obtained from Hall measurements.

The devices were fabricated by the following process. First, a Bi$_2$Se$_3$ film is etched using Ar ion milling to the desired thickness, and is patterned into a 800 $\mu$m × 600 $\mu$m film using photolithography and Ar ion milling. Then, a 20 nm Py (Ni$_{81}$Fe$_{19}$) layer is sputter deposited and patterned into a 640 $\mu$m × 600 $\mu$m strip. In the next step, a 600 $\mu$m × 630 $\mu$m × 30 nm SiO$_2$ layer is deposited to isolate the Py layer. Then, in the last step, Ta (5 nm)/Cu (150 nm) asymmetric coplanar strips (ACPS) and dc probe pads are patterned, and sputter deposited, simultaneously. In the ACPS, the width of signal line is 60 $\mu$m, the width of ground line is 180 $\mu$m, and the signal-ground spacing is 30 $\mu$m.

Figure 1(a) shows the resistivity and the carrier concentration of a 20 quintuple layer (QL; 1 QL≈1 nm) Bi$_2$Se$_3$ film as a function of temperature. The decrease in the resistance with decreasing temperature is a typical characteristic of Bi$_2$Se$_3$ and the surface dominant transport in these films has been previously confirmed by sample thickness dependent studies of transport properties [19, 20]. Figure 1(b) elucidates the spin pumping process. The precessing magnetization $M$ of the ferromagnet gives rise to a spin current $J_s$ from the ferromagnet to a non-magnetic material (Bi$_2$Se$_3$). This spin current is detected as a transverse voltage via ISHE in Bi$_2$Se$_3$, given by $V_{\text{ISHE}} \propto J_S \times \sigma$, where $\sigma$ denotes the spin polarization vector of the spin current. Figure 1(c) shows a schematic of the device with the measurement set-up. The cross-section of the device is shown in Fig. 1(d). For spin pumping induced ISHE measurements, a 15 dBm microwave signal of a fixed frequency of 3, 4, 5, and 6 GHz ($=\omega/2\pi$) is applied to the asymmetric coplanar strips (ACPS) waveguide using a signal generator (SG) and a dc voltage is measured across the Bi$_2$Se$_3$/Py bilayers as a function of the bias field ($H_b$) applied along the $z$-direction. For the evaluation of enhancement in the Gilbert damping constant ($\alpha$), ferromagnetic resonance (FMR) measurements are carried out using a vector network analyzer. Measurements are carried out at different temperatures ranging from 15 to 300 K.

Figure 2(a) shows the measured data of the electromotive force signal in a Bi$_2$Se$_3$ (20 nm)/Py device measured at a frequency of 4 GHz. The $H_b$ along the $z$-direction is swept across the resonance field $H_0$, which is related to the applied frequency by Kittel formula, $\omega = \gamma\mu_0\sqrt{H_0(H_0+M)}$, where $\gamma$ is the gyromagnetic ratio of a free electron, $\mu_0$ is the permeability of free space, and $M$ is the saturation magnetization of Py. The measured signal originates from three sources; the ISHE in the Bi$_2$Se$_3$ layer, the anisotropic magnetoresistance (AMR) of the Py layer, and the anomalous Hall effect (AHE) of the Py layer [6, 35, 36]. The effect of AMR or AHE from the Py layer needs to be eliminated considering that the ISHE signal has a symmetric Lorentzian shape, whereas the signal due to AMR or AHE has an asymmetric Lorentzian shape [6, 35]. Therefore, the measured data are fitted by a sum of symmetric and asymmetric Lorentzian functions, $V = V_{sym}\dfrac{\Gamma^2}{\Gamma^2+(H-H_0)^2} + V_{asym}\dfrac{\Gamma(H-H_0)}{\Gamma^2+(H-H_0)^2}$ and the value of $V_{sym}$ is taken to be the spin pumping induced ISHE signal ($V_{ISHE}$). It must be noted that spin rectification effects may also contribute to the symmetric part of the signal [37-39]. These effects originate from the AMR of the Py layer, therefore in order to rule out such a contribution, we perform measurements on a bare Py layer (without Bi$_2$Se$_3$). Only a small asymmetric signal but no symmetric signal is observed as shown in Fig. 2(a).

The spin current density induced by spin pumping is given by [40]

$$J_s = \frac{\hbar g_{r\uparrow\downarrow} \gamma^2 h_{rf}^2 \left(M\gamma + \sqrt{M^2\gamma^2 + 4\omega^2}\right)}{8\pi\alpha^2 \left(M^2\gamma^2 + 4\omega^2\right)} \frac{\lambda_{sf}}{d_{BiSe}} \tanh\left(\frac{d_{BiSe}}{2\lambda_{sf}}\right) \quad (1)$$

where $h_{rf}$ is the *rf* field amplitude, $f$ is the applied frequency ($\omega = 2\pi f$), $\lambda_{sf}$ is the spin diffusion length in Bi$_2$Se$_3$, and $d_{BiSe}$ is the thickness of the Bi$_2$Se$_3$ layer. $g_{r\uparrow\downarrow}$ is the spin mixing conductance of the Bi$_2$Se$_3$/Py interface and is a measure of efficiency of spin pumping across the interface [41, 42]. Due to the ISHE in the Bi$_2$Se$_3$ layer, this spin current is converted into a charge current with a density given by $J_c = \theta_{sh}(2e/\hbar)J_s$, where $\theta_{sh}$ is the spin Hall angle in Bi$_2$Se$_3$. For the device width $w$ (see Fig. 1(c)), this charge current $I_c = J_c(wd_{BiSe})$ across the resistance $R$ of the Bi$_2$Se$_3$/Py bilayer is detected as an emf, $V_{ISHE} = J_c(wd_{BiSe})R$. Thus the measured ISHE signal can be written as

$$\frac{V_{ISHE}}{R} = \theta_{sh} w d_{BiSe} \left(\frac{2e}{\hbar}\right) \frac{\hbar g_{r\uparrow\downarrow} \gamma^2 h_{rf}^2 \left(M\gamma + \sqrt{M^2\gamma^2 + 4\omega^2}\right)}{8\pi\alpha^2 \left(M^2\gamma^2 + 4\omega^2\right)} \frac{\lambda_{sf}}{d_{BiSe}} \tanh\left(\frac{d_{BiSe}}{2\lambda_{sf}}\right). \quad (2)$$

The value of $h_{rf}$ can be estimated used microwave photoresistance measurements [43-45]. Figure 2(b) shows the microwave photoresistance data from a Bi$_2$Se$_3$ (20 nm)/Py device at 4 GHz. From these measurements, the cone angle of precession is calculated to be 0.23°, and $h_{rf}$ is estimated to be ~ 1 Oe using the method described in a previous report [44]. It is also noteworthy that the photoresistance signals in our devices are of the same magnitude in both positive and negative $H_b$, which is in agreement with the equal magnitude of spin pumping induced ISHE signals at ± $H_b$.

In order to determine the spin Hall angle and spin diffusion length in the Bi$_2$Se$_3$ layer, spin pumping induced ISHE measurements were carried out in samples with different thicknesses of the Bi$_2$Se$_3$ layer. Figure 2(c) shows the measured value of $V_{ISHE}/R$ for different thicknesses of the Bi$_2$Se$_3$ layer, and a corresponding fit by Eq. (2). From the fitting we obtain a value of $\lambda_{sf} = 6.22 \pm 0.15$ nm at room temperature for the spin diffusion length. The fitting also gives us the product $g_{r\uparrow\downarrow}\theta_{sh} = (4.57 \pm 0.12) \times 10^{16}$ m$^{-2}$, where $g_{r\uparrow\downarrow}$ can be obtained by using ferromagnetic resonance (FMR) measurements. Since spin pumping is a process in which the precessing magnetization loses energy, the effective Gilbert damping increases due to spin pumping [41, 46]. In Fig. 2(d) the FMR full linewidth measured as a function of frequency is seen to be larger in Bi$_2$Se$_3$/Py devices than that in Py devices. The values of $\alpha$ for Py and Bi$_2$Se$_3$/Py from the linear fits by the equation $\Delta H = \Delta H_0 + 4\pi\alpha f/\gamma$ are found to be 0.0109 and 0.0122, respectively. The spin mixing conductance is related to the enhancement in effective $\alpha$ via $g_{r\uparrow\downarrow} = 4\pi M d_{Py}(\Delta\alpha)/(g\mu_B)$. [35, 40] Thus, $g_{r\uparrow\downarrow}$ is found to be $1.514 \times 10^{19}$ m$^{-2}$ and $\theta_{sh}$ to be $0.0093 \pm 0.0013$ at room temperature.

It must be noted that recently a very large value of $\theta_{sh}$ (2 – 3.5) was reported in Bi$_2$Se$_3$ by spin torque ferromagnetic resonance (ST-FMR) measurements in Bi$_2$Se$_3$/Py devices [26]. In another similar material, (Bi$_{0.5}$Sb$_{0.5}$)$_2$Te$_3$, an even larger $\theta_{sh}$ (140 – 425) was reported using spin-orbit switching measurements in (Bi$_{0.5}$Sb$_{0.5}$)$_2$Te$_3$/(Cr$_{0.08}$Bi$_{0.54}$Sb$_{0.38}$)$_2$Te$_3$ heterostructures [27]. Our value of $\theta_{sh}$ in Bi$_2$Se$_3$ is much smaller and closer to that in a conventional heavy metal like Pt [30, 47]. One possibility for the discrepancy could be that both ST-FMR and spin-orbit switching measurements involve a current injection through the TI material, in which the spin-momentum locking mechanism may play an important role in determining $\theta_{sh}$, unlike our spin pumping experiments. It is also possible that the quality of TI films plays a role in the above discrepancy as it is known that TI materials pose challenges in high quality film growth.

We further extended our study to low temperatures. The $Bi_2Se_3$ thickness dependent measurements of $V_{ISHE}$ were carried out at different temperatures ranging from 15 to 300 K in order to obtain the $\theta_{sh}$ and $\lambda_{sf}$ at each temperature. Figure 3(a) shows the variation of $\theta_{sh}$ and $\lambda_{sf}$ as a function of temperature. Both $\theta_{sh}$ and $\lambda_{sf}$ increase at low temperatures to be $\theta_{sh} = 0.022 \pm 0.0028$ and $\lambda_{sf} = 9.5 \pm 0.35$ nm at 15 K. The increase in $\lambda_{sf}$ may be related to a decrease in phonon scattering or an increase in the carrier mobility as temperature is lowered. $\lambda_{sf}$ is related to the spin diffusion time ($\tau_{sf}$) via $\lambda_{sf} = \sqrt{D\tau_{sf}}$, where $D$ is the diffusion constant. $D$ is related to the carrier mobility through Einstein's relation $D = \mu k_B T / e$. The carrier mobility in our $Bi_2Se_3$ film is found to increase by three times at low temperature [48], thus qualitatively explaining the increase in $\lambda_{sf}$ at low temperatures.

The spin orbit length ($l_{so}$) in $Bi_2Se_3$ was obtained by weak antilocalization measurements. For these measurements the 100 $\mu$m × 20 $\mu$m films were fabricated by photolithography by Ar ion milling process and the electrical contacts, Ta (5 nm)/Cu (80 nm), were sputter deposited. The electrical resistance of the $Bi_2Se_3$ films was measured as a function of an out-of-plane magnetic field. We fit this data with the Hikami-Larkin-Nagaoka (HLN) equation along with a quadratic background term [49]

$$\Delta G(B) = -\frac{Ae^2}{\pi h}\left[\psi\left(\frac{\hbar}{4eL^2B} + \frac{1}{2}\right) - \ln\left(\frac{\hbar}{4eL^2B}\right)\right] + \beta B^2 \qquad (3)$$

Here, *A* accounts for contributions from the surface as well as 2D bulk effects, and $\beta$ is the quadratic coefficient arising from scattering events. $\psi$ is the digamma function, *L* is the phase coherence length in $Bi_2Se_3$, *e* is the electronic charge, *h* is Planck constant, and $\hbar$ is the reduced Planck constant. Figure 3(b) shows the measured magnetoconductance (symbols) and the corresponding fit (solid line) by equation (3). From fitting we obtain the value of $\beta$ to be $-4.32\times10^{-9}$ $\Omega^{-1}T^{-2}$. The $\beta$ consists of a classical cyclotronic part, and a quantum part that originates from spin orbit scattering and elastic scattering events. The classical part is given by $\beta_c = -\mu_H^2 G_0$, where $\mu_H$ is the Hall mobility and $G_0$ is the zero-field conductance. The quantum part of $\beta$ is given by

$$\beta_q = -\frac{e^2}{24\pi h}\left[\frac{1}{B_{so}+B_e}\right]^2 + \frac{3e^2}{48\pi h}\left[\frac{1}{(4/3)B_{so}+B_\phi}\right]^2 \qquad (4)$$

where $B_{so} = \hbar/(4el_{so}^2)$ and $B_e = \hbar/(4el_e^2)$, with $l_{so}$ and $l_e$ being the spin orbit length and the electron mean free path, respectively. The mobility in our samples was determined to be 61 cm$^2$/V·sec, and $G_0$ is $5.43\times10^{-4}$ $\Omega^{-1}$. Thus the value of $\beta_c$ is calculated to be $-2.02\times10^{-8}$ $\Omega^{-1}T^{-2}$. Thus, $\beta_q = \beta - \beta_c = 1.58\times10^{-8}$ $\Omega^{-1}T^{-2}$. If we use the electron mean free path, $l_e = 10$ nm, the $l_{so}$ can be obtained to be 6.9 nm. The comparable values of $l_{so}$ and $\lambda_{sf}$ suggest that spin-orbit coupling is the dominant source of spin scattering in the $Bi_2Se_3$ films.

$Bi_2Se_3$ films are expected to show a distinct behavior at surface and bulk. For example, the spin-orbit interaction at surface may be different from that in bulk. We thus analyze our data using different values of spin Hall angle for the surface and bulk. Following the method by Ando *et al* [40]. we derive the formula for spin pumping induced $V_{ISHE}$ using different values of spin Hall angle for the surface and bulk,

$$\begin{aligned}
\frac{V_{ISHE}}{R} &= \theta_{sh1}\left(\frac{2e}{\hbar}\right)\frac{\lambda_{sf}}{d_{BiSe}}\operatorname{csch}\left(\frac{d_{BiSe}}{\lambda_{sf}}\right)\left(\cosh\left(\frac{d_{BiSe}}{\lambda_{sf}}\right)-\cosh\left(\frac{d_{BiSe}-d_{surf}}{\lambda_{sf}}\right)\right)J_s^0 \\
&+ \theta_{sh2}\left(\frac{2e}{\hbar}\right)\frac{\lambda_{sf}}{d_{BiSe}}\operatorname{csch}\left(\frac{d_{BiSe}}{\lambda_{sf}}\right)\left(\cosh\left(\frac{d_{BiSe}-d_{surf}}{\lambda_{sf}}\right)-\cosh\left(\frac{d_{surf}}{\lambda_{sf}}\right)\right)J_s^0 \\
&+ \theta_{sh1}\left(\frac{2e}{\hbar}\right)\frac{\lambda_{sf}}{d_{BiSe}}\operatorname{csch}\left(\frac{d_{BiSe}}{\lambda_{sf}}\right)\left(1-\cosh\left(\frac{d_{surf}}{\lambda_{sf}}\right)\right)J_s^0, \quad J_s = J_s^0 \frac{\lambda_{sf}}{d_{BiSe}}\tanh\left(\frac{d_{BiSe}}{2\lambda_{sf}}\right)
\end{aligned}$$

(5)

where $\theta_{sh1}$ and $\theta_{sh2}$ are the spin Hall angles at surface and bulk, respectively. $d_{surf}$ is the assumed value of surface thickness and is taken to be 3 nm. The spin Hall angle at opposite surfaces was taken to be of opposite signs. Fitting the measured data, we obtain $\theta_{sh1}$ at surface and $\theta_{sh2}$ at bulk for each temperature as shown in Fig. 4(a), which does not show any clear distinction between the surface and bulk value. Figure 4(b) shows the $\lambda_{sf}$ as a function of temperature when different values of spin Hall angle are taken for surface and for bulk.

Our results show the practicability of $Bi_2Se_3$ as a spin detector. The value of $\theta_{sh}$ is higher than that in Si and GaAs, and similar to those obtained for heavy metals such as Pt, Pd, or Ta by spin pumping method [32, 33, 36, 40]. It must be noted that the spin current detection efficiency by ISHE depends not only on $\theta_{sh}$, but also on the resistance of the layer, as $V_{ISHE} \propto \theta_{sh} J_s R$. Since $Bi_2Se_3$ is a semiconductor, its resistivity is at least an order of magnitude higher than the heavy metals like Pt and hence the spin pumping induced ISHE voltage can be expected to be higher than that in the Pt [50]. Therefore, $Bi_2Se_3$ can be a viable material for spin detection devices.

This work is partially supported by the Singapore Ministry of Education Academic Research Fund Tier 1 (R-263-000-A75-750), the National Research Foundation, Prime Minister's Office, Singapore under its Competitive Research Programme (CRP Award No. NRF-CRP 4-2008-06), National Science Foundation (NSF DMR-0845464), and Office of Naval Research (ONR N000141210456).

*eleyang@nus.edu.sg

**Figure captions**

Fig. 1. (a) Resistivity and carrier concentration of 20 QL $Bi_2Se_3$ as a function of temperature. (b) A schematic representation of spin pumping. (c) An illustration of the measurement setup. A rectangular $Bi_2Se_3$/Py layer is patterned. An ACPS line is patterned on top of the $SiO_2$ layer. (d) Cross-section of the device. The layer stack is $Bi_2Se_3$ ($t$)/$Ni_{81}Fe_{19}$ (20 nm)/$SiO_2$ (30 nm), with $t = 5 - 35$ nm. The electrical contacts are Ta (5 nm)/Cu (150 nm).

Fig. 2. (a) Spin pumping induced ISHE signal ($V_{ISHE}$) measured at 4 GHz from the device with a $Bi_2Se_3$ thickness ($t$) of 20 nm, and in a bare Py (20 nm) film. The measured ISHE data (symbols) is fitted with a sum of a symmetric and an asymmetric Lorentzian. (b) Microwave photoresistance signals ($\Delta R_{MW}$) at 4 GHz in the device with a $Bi_2Se_3$ thickness ($t$) of 20 nm. (c) The ratio of $V_{ISHE}$ and resistance ($R$) of $Bi_2Se_3$/Py films. (d) The FMR linewidth as a function of frequency for the Py and $Bi_2Se_3$/Py films.

Fig. 3. (a) The spin Hall angle ($\theta_{sh}$) and spin diffusion length ($\lambda_{sf}$) in $Bi_2Se_3$ as a function of temperature. (b) The measured magnetoconductance data (symbols) and corresponding fit (line) using Hikami-Larkin-Nagaoka (HLN) equation.

Fig. 4. (a) The spin Hall angle of $Bi_2Se_3$ surface and $Bi_2Se_3$ bulk, assuming a surface thickness of 3 nm, as a function of temperature. (b) The spin diffusion length of $Bi_2Se_3$ assuming different spin Hall angles in surface and bulk, as a function of temperature.

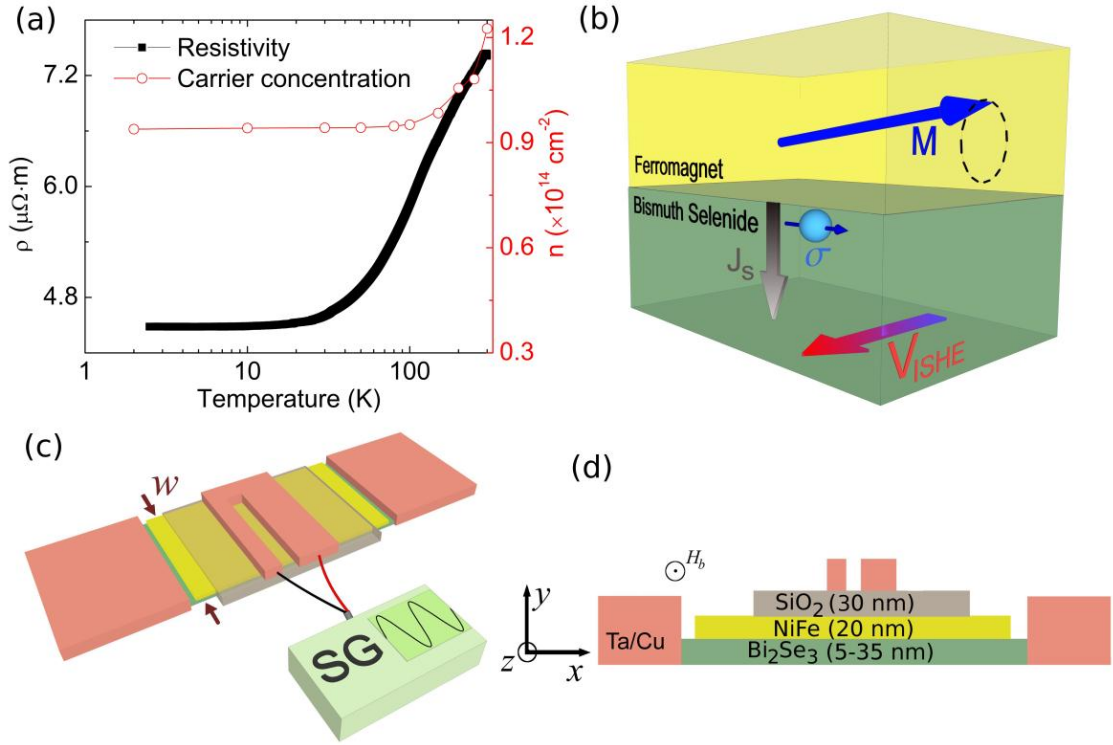

Fig. 1

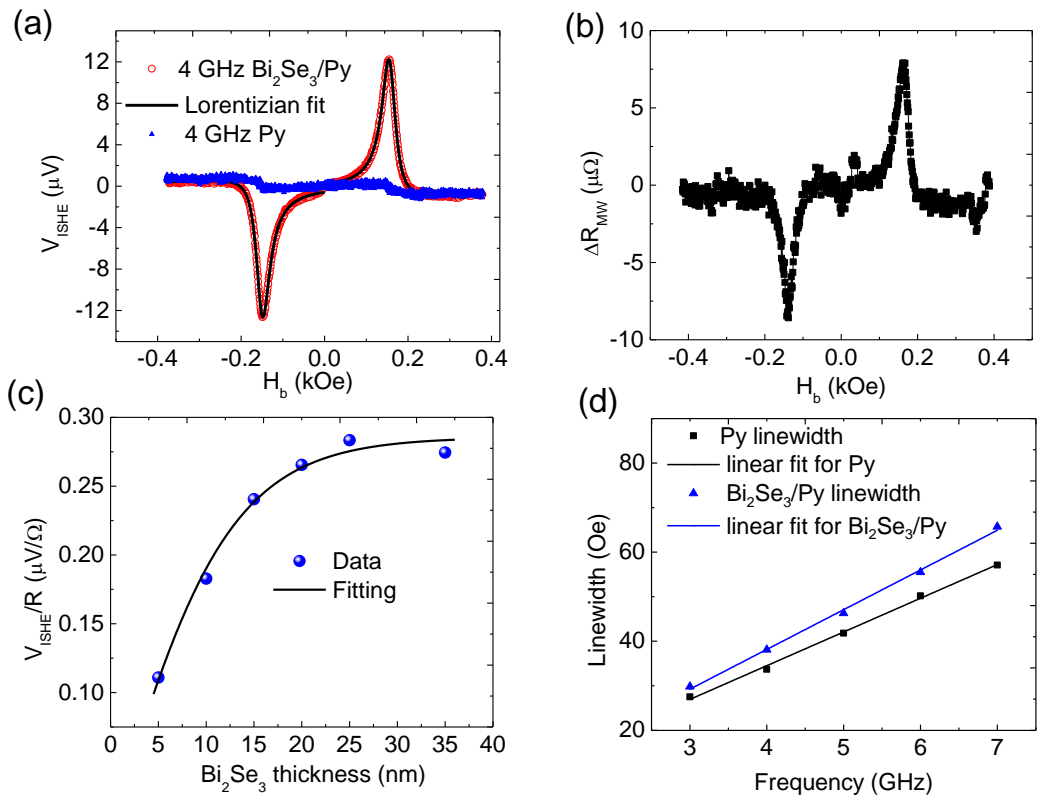

Fig. 2

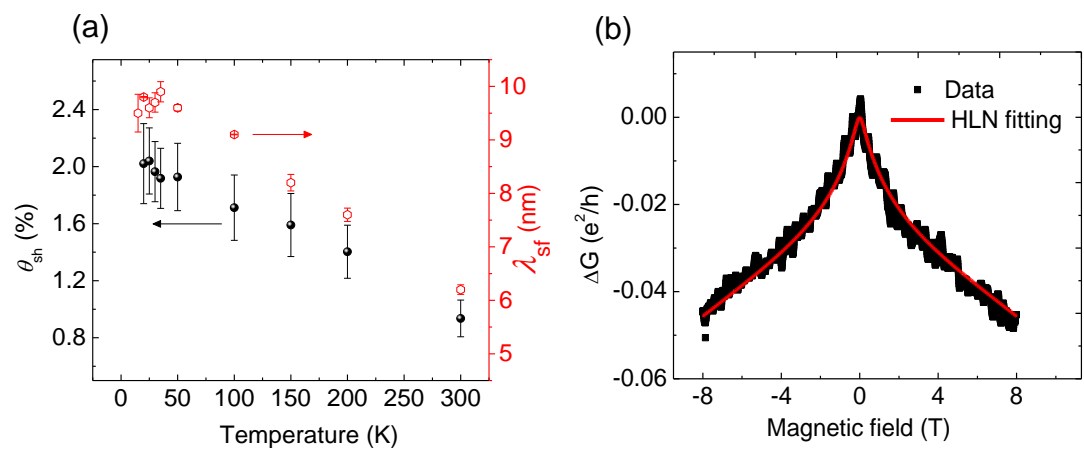

Fig. 3

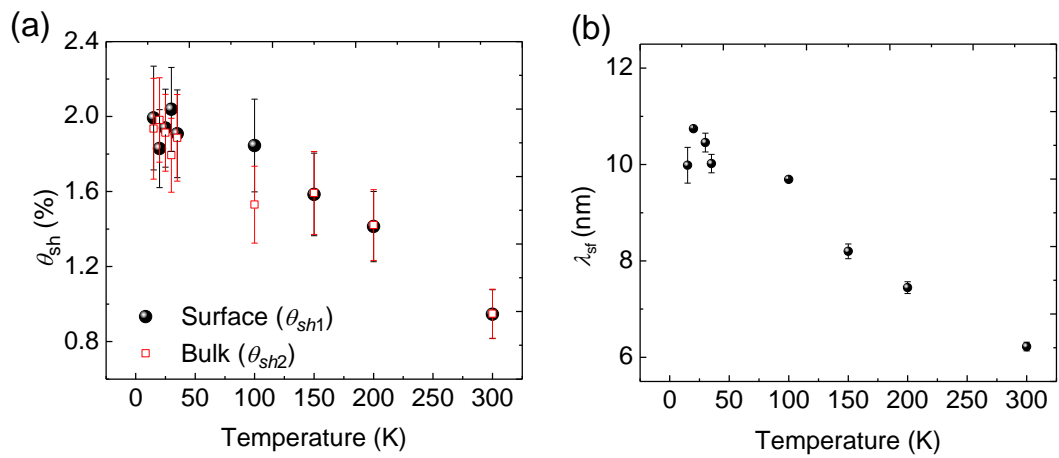

Fig. 4